\documentclass[letter]{aa}
\usepackage{amssymb,amsmath}
\usepackage{graphicx}
\usepackage{xcolor,natbib}
\usepackage{multirow}
\usepackage[T1]{fontenc}
\usepackage{ae,aecompl}
\usepackage{newtxtext, newtxmath}
\usepackage{float}
\usepackage{subfigure}
\usepackage{longtable}
\usepackage{enumitem}
\usepackage{longtable,listings}
\usepackage[flushleft]{threeparttable}
\usepackage{parcolumns}
\usepackage{hyperref}

\def\n2{[N~{\sc ii}]$\lambda6583$\AA}

\def\o3{[O~{\sc iii}]$\lambda5007$\AA}
\def\obj{SDSS J0012-1022}

\begin{document}

\title{Variability of optical spectral index to support a central sub-parsec binary black 
hole system in quasar SDSS J001224-102226.51}

\titlerunning{variations in optical spectral index}


\author{XueGuang Zhang}

\institute{Guangxi Key Laboratory for Relativistic Astrophysics, School of Physical Science 
and Technology, GuangXi University, 
No. 100, Daxue Road, Nanning, 530004, P. R. China \ \ \ \email{xgzhang@gxu.edu.cn}}

\abstract
{In this manuscript, variations in optical spectral index $\alpha_{5100}$ are applied for 
detecting central sub-parsec binary black hole systems (sub-pc BBHs) in broad line active 
galactic nuclei (BLAGN), due to apparent effects of obscurations on central two BH accreting 
systems. For sub-pc BBHs in BLAGN, two main characteristics on $\alpha_{5100}$ can be expected. 
First, if a BLAGN harbours a central sub-pc BBH, the expected unique variability in 
$\alpha_{5100}$ should lead the BLAGN to be an outlier in the space of $\alpha_{5100}$ versus 
continuum luminosity $L_{5100}$ determined from normal BLAGN. Second, BLAGN harbouring central 
sub-pc BBHs could lead to periodic variations in $\alpha_{5100}$. Here, after checking the 
two-epoch SDSS spectra of quasar SDSS J0012-1022 reported as a candidate of sub-pc BBH by 
large velocity offset between narrow and broad Balmer emission lines, unique variability of 
$\alpha_{5100}$ can be explained by effects of obscurations related to an assumed central 
sub-pc BBH. In the near future, to detect and report periodic variations of $\alpha_{5100}$ 
for sub-pc BBHs in BLAGN should be our main objective. The results provide a new method by 
applications of properties of optical continuum emissions for detecting sub-pc BBHs in BLAGN.}


\maketitle

\section{Introduction}

	Sub-parsec binary black hole systems (sub-pc BBHs) in broad line active galactic 
nuclei (BLAGN) have been accepted as main targets for gravitational wave signals at nano-Hz 
frequencies \citep{fb90, ar15, sh18, is22, rz23, pt24, sr25, ti25}, to provide further clues 
on final stages of BH merging as discussed in \citet{bb80, mk10, fg19, mj22, ak24}. Therefore, 
it is necessary and meaningful to detect more candidates of sub-pc BBHs.  

	Different methods have been proposed for detecting sub-pc BBHs in BLAGN through 
properties of photometric variability and spectroscopic features. Considering long-standing 
optical Quasi-Periodic Oscillations (QPOs) with periodicities around hundreds to thousands 
of days related to orbital motions of sub-pc BBHs, optical QPOs detected in photometric 
variability have been accepted as indicators for sub-pc BBHs, as reported in \citet{gd15a, 
gd15b, cb16, zb16, zh22a, zh22b, zh23a, zh25, zh25c}. Meanwhile, unique properties of 
spectroscopic emission line features have been applied to detect sub-pc BBHs. \citet{eb12} 
have reported a sample of sub-pc BBH candidates through broad H$\beta$ emission lines 
displaced from narrow emission lines with shifted velocities larger than 1000km/s. 
\citet{bl09} have reported the sub-pc BBH candidate in SDSS J1536+0441 through unique 
double-peaked features in broad emission lines. \citet{zh21a, zh23b, zh25b} have reported 
the sub-pc BBH candidates in SDSS J1547, SDSS J1257 and PG 1411+442 through different line 
profiles of broad Balmer emission lines.

	Besides expected unique spectroscopic emission line properties, the intrinsic 
dependence of broad line luminosity on continuum luminosity \citep{gh05} should indicate 
spectroscopic continuum emissions (such as spectral index) in multi-epoch spectra could 
also provide efficient clues to support sub-pc BBHs in BLAGN. Therefore, to detect and 
report such clues is the main objective of the manuscript. Here, only the sub-pc BBHs with 
separation large enough are mainly considered, with little contribution of 
circum-binary disk \citep{nm12, dh15, bc17, nk21, ak24} to the optical continuum emissions.

	In this manuscript, Section 2 presents the spectroscopic results and main discussions 
of the BLAGN SDSS J001224-102226.51 (=SDSS J0012-1022) at redshift 0.228 reported in 
\citet{eb12}, showing unique variability properties of optical spectral index, providing 
further clues to support a central sub-pc BBH. Section 3 shows necessary discussions on optical 
QPOs in $gr$-band light curves from Zwicky Transient Facility (ZTF) \citep{bk19, ml19} in 
\obj. Main conclusions are given in Section 4. Throughout the manuscript, we have adopted 
the cosmological parameters of $H_{0}$=70 km s$^{-1}$ Mpc$^{-1}$, $\Omega_{m}$=0.3, and 
$\Omega_{\Lambda}$=0.7.

\begin{figure}
\centering\includegraphics[width = 9cm,height=11cm]{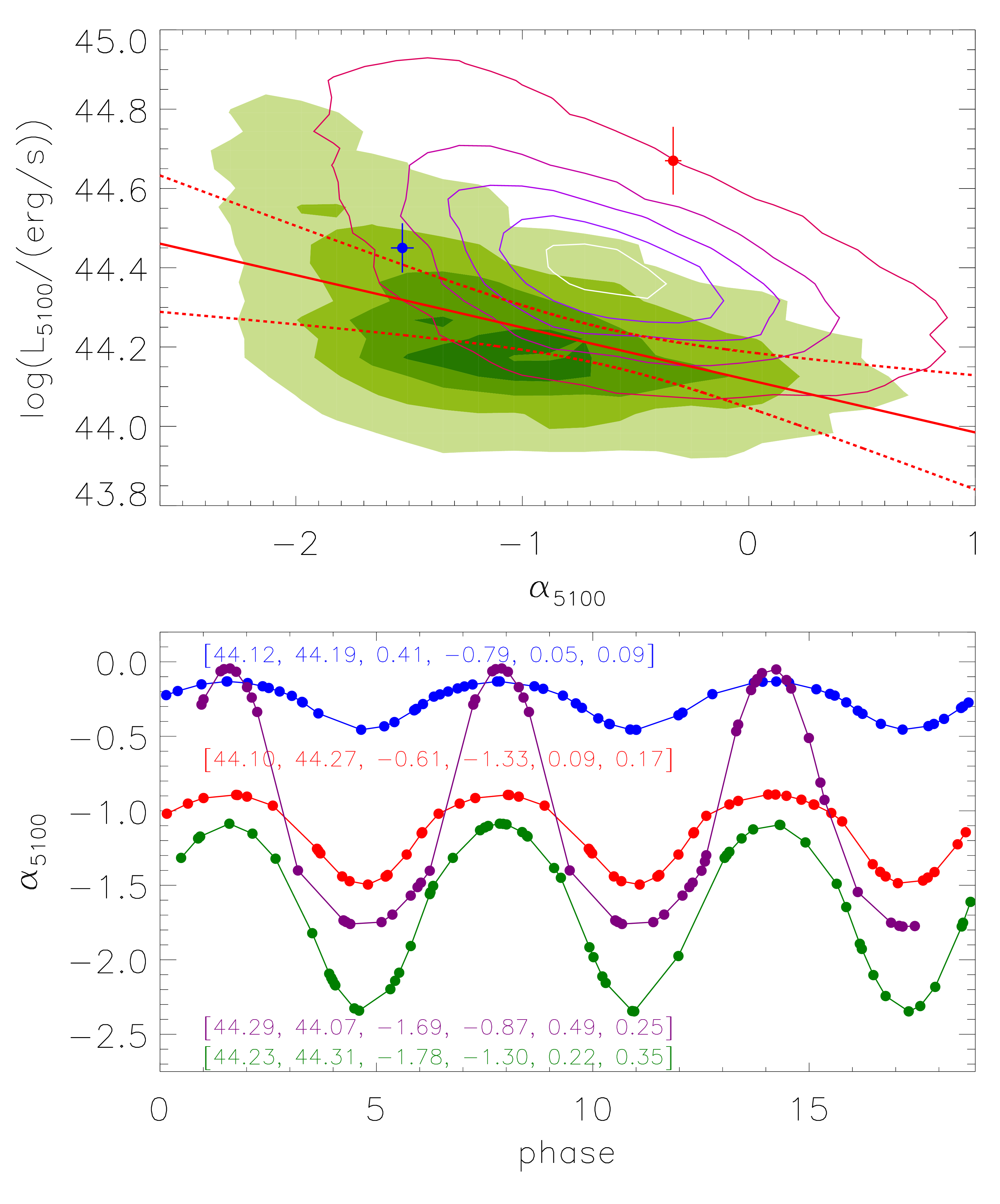}
\caption{Top panel shows the contour (filled by greenish colors) for the correlation between 
$L_{5100}$ and $\alpha_{5100}$ of the selected 3274 SDSS quasars. Circles plus error bars in 
blue and in red show the results of SDSS J0012-1022 with MJD=52141 and 56628. Solid and dashed 
lines in red show the best fitting results and corresponding 5$\sigma$ confidence bands. 
Contour with reddish levels represents the results through the 20000 artificial $f_{\lambda}$. 
The contour levels in each contour represent 0.1, 0.3, 0.5, 0.7 and 0.9 of the 2D volume 
contained. Bottom panel shows four examples on periodic variations in $\alpha_{5100}$ related 
to sub-pc BBHs, with corresponding model parameters listed in the same color. The six parameters 
in each square bracket are $\log(L_1)$, $\log(L_2)$, $\alpha_1$, $\alpha_2$, $E_{10}$ and 
$E_{20}$.}
\label{sed}
\end{figure}

\section{Spectroscopic results for SDSS J0012-1022} 

	Considering normal BLAGN, there is an apparent dependence of optical spectral index 
$\alpha_{5100}$ ($f_\lambda\propto\lambda^{\alpha_{5100}}$) underneath broad H$\beta$ on 
continuum luminosity at 5100\AA~ $L_{5100}$, as shown in top panel of Fig.~\ref{sed} for the 
selected 3274 low redshift ($z<0.35$) SDSS quasars in \citet{sr11} with reliable measurements 
of $\alpha_{5100}$ (between -5 and 3) and $L_{5100}$ (5 times larger than corresponding 
uncertainties), with Spearman rank correlation coefficient -0.51 ($P_{null}<10^{-10}$). After 
considering uncertainties in both coordinates, through the Least Trimmed Squares regression 
technique \citep{cap13}, the dependence can be described as 
$\log{\frac{L_{5100}}{\rm erg/s}}=(44.117\pm0.005)-(0.132\pm0.004)\alpha_{5100}$ with RMS 
scatter of 0.149. The results are consistent with results in \citet{pc94, sr12, pp25} that 
vast majority of BLAGN get bluer when they get brighter.

	Once assumed a sub-pc BBH in a BLAGN, different obscurations on continuum emissions 
related to two BH accreting systems could be naturally expected, and inevitably have a 
strong dependence on inclination angle. If one BLAGN harbouring a sub-pc BBH should have 
unique variability in optical spectral index, leading the BLAGN to be an outlier in the 
space of $\alpha_{5100}$ versus $L_{5100}$ for the normal quasars, such as the following 
discussed results in \obj. 

\begin{figure}
\centering\includegraphics[width = 9cm,height=5cm]{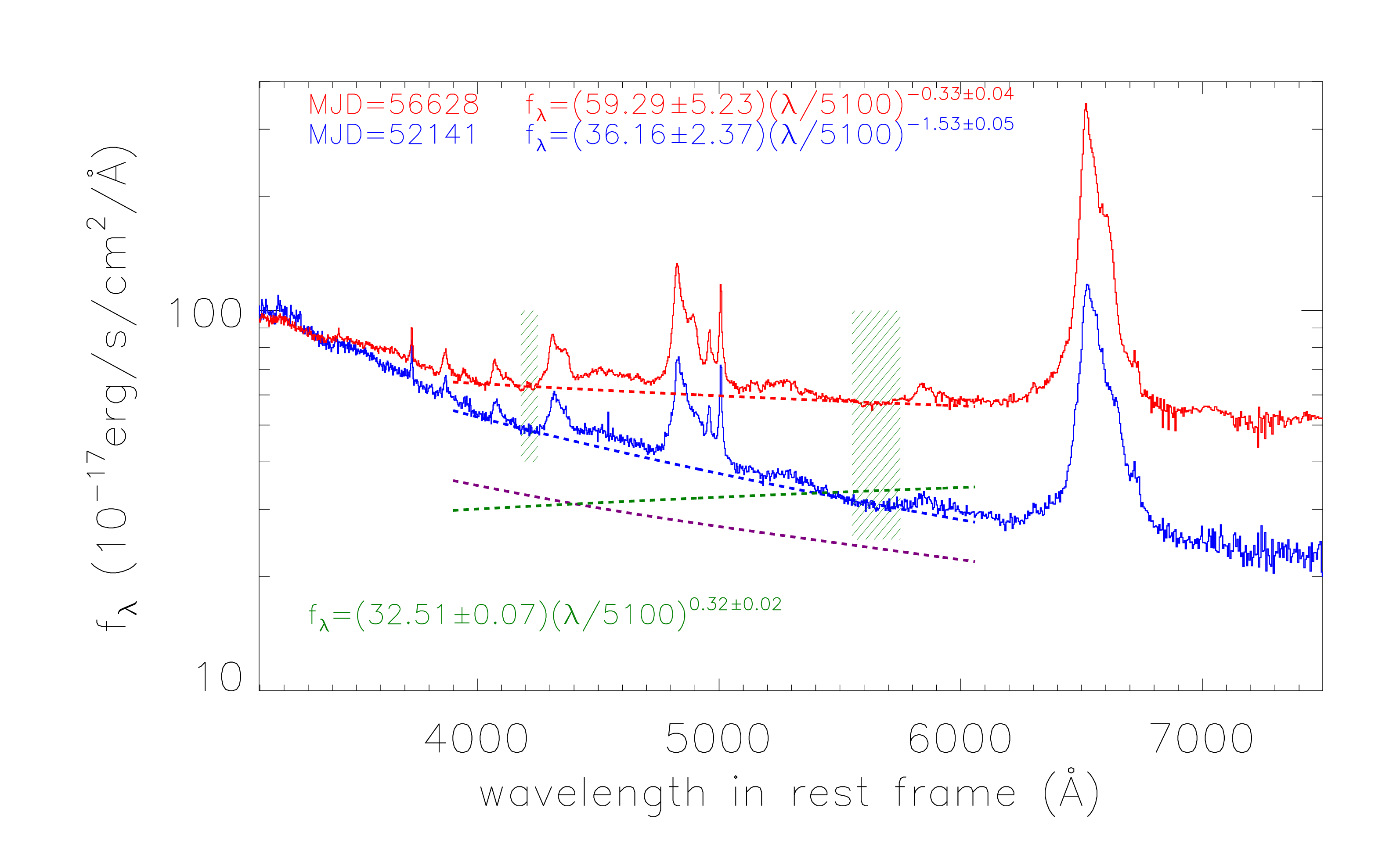}
\caption{SDSS spectra of SDSS J0012-1022 with MJD = 52141 (in blue) and 56628 (in red). Dashed 
lines in blue and in red show the power law function determined optical continuum emissions 
underneath broad H$\beta$ through the two regions filled by dark green lines. Dashed purple 
line represents the first component related to the obscured power law component shown as 
dashed blue line with E(B-V)=0.1, and dashed dark green line shows the second component 
(formula listed in dark green characters) from the other one BH accreting system. The sum of 
the two components is one solution related to a sub-pc BBH to the continuum emissions in the 
spectrum with MJD=56628.}
\label{spec}
\end{figure}

	\obj~ is a quasar at redshift 0.228\ in Sloan Digital Sky Survey (SDSS) \citep{al23}, 
with its high-quality SDSS spectra (plate-mjd-fiberid=0651-52141-0072 and 7169-56628-0344) 
observed at MJD=52141 and 56628 shown in Fig.~\ref{spec}. The power law continuum emissions 
$f_\lambda=A(\frac{\lambda}{5100})^{\alpha_{5100}}$ underneath broad H$\beta$ can be determined 
by the continuum emissions with rest wavelength from 4180 to 4250\AA~ and from 5550 to 5750\AA~ 
through the Levenberg-Marquardt least-squares minimization technique \citep{mc09}, leading to 
$A=36.16\pm2.37$ $\alpha_{5100}=-1.53\pm0.05$ and $A=59.29\pm5.23$, $\alpha_{5100}=-0.33\pm0.04$ 
for the spectrum with MJD=52141 and 56628, respectively.
In top panel of Fig.~\ref{sed}, the determined $\alpha_{5100}$ and $L_{5100}$ lead \obj~ with 
MJD=52141 (solid blue circle) to lie in the common region in the space of $\alpha_{5100}$ 
versus $L_{5100}$ for normal quasars, however, the determined values with MJD=56628 apparently 
lead \obj~ (solid red circle) to be an outlier having opposite behaviour in the space. In 
other words, when \obj~ got brighter from MJD=52141 to 56628, there was not bluer but redder 
continuum emissions around 5100\AA. 

	Based on the results in top panel of Fig.~\ref{sed} for \obj, obscurations on optical 
continuum emissions in normal BLAGN cannot be accepted. Then, we check whether an assumed 
sub-pc BBH can be applied.

	For an assumed sub-pc BBH, observed optical continuum emissions $f(\lambda)$ 
($\lambda$ from 4400\AA~ to 5600\AA~ in rest frame) actually include two sources of 
$f_1(\lambda, ob)$ and $f_2(\lambda, ob)$ with different spectral indices from central 
two BH accreting systems,
\begin{equation}
f(\lambda)=f_1(\lambda, ob)+f_2(\lambda,ob)=k_1(\lambda,\phi) f_1(\lambda)
	+k_2(\lambda,\phi) f_2(\lambda)
\end{equation},
with $k_1(\lambda,\phi)$ and $k_2(\lambda,\phi)$ ($\phi$ as orbital phase) representing effects 
of obscurations on intrinsic $f_1(\lambda)\propto(\lambda/5100)^{\alpha_1}$ (continuum luminosity 
$L_1$ at 5100\AA) and $f_2(\lambda)\propto(\lambda/5100)^{\alpha_2}$ (continuum luminosity 
$L_2$ at 5100\AA).

	Meanwhile, $k_1(\lambda,\phi)$ and $k_2(\lambda,\phi)$ can be simply determined by 
evolutions of color excess E(B-V) simply described by the following periodic functions
\begin{equation}
\begin{split}
	E_1(B-V,\phi)~&=~E_{10}\times |sin(\phi)| \ \ (\phi\in[\phi_1,~\phi_2]) \\
	E_1(B-V,\phi)~&=~0  \ \ (\phi\in[0,~\phi_1] \ \ or \ [\phi_2,~2\pi]) \\
	E_2(B-V,\phi)~&=~E_{20}\times |sin(\phi)| \ \ (\phi\in[0,~\phi_1] \ \ or \ [\phi_2,~2\pi]) \\
	E_2(B-V,\phi)~&=~0  \ \ (\phi\in[\phi_1,~\phi_2])
\end{split}.
\end{equation}
The parameters of $\phi_1$ and $\phi_2$ ($0~\le~(\phi_2 - \phi_1)~\le~\pi$) are applied to 
determine the phase information for existence of covered regions of one BH accreting system 
by the other system. And, the extinction curve in \citet{fi99} has been accepted. Then, 
assumed sub-pc BBHs, artificial $f(\lambda)$ can be created by the following steps.

	First, among the 3274 SDSS quasars shown in top panel of Fig.~\ref{sed}, two quasars 
are random selected, leading to the measured values of the two quasars as the intrinsic $L_1$, 
$\alpha_1$ and $L_2$, $\alpha_2$. Second, according to the redshift of one of the two selected 
quasars in the first step, $f_1(\lambda)$ and $f_2(\lambda)$ can be described as 
\begin{equation}
f_i(\lambda)=\frac{L_i}{4\pi\times5100\times D^2}(\frac{\lambda}{5100})^{\alpha_i} \ \ \ (i=1, 2)
\end{equation}
with $D$ as the redshift determined luminosity distance. Third, $\phi_1$ and 
$2\pi>\phi_2>\phi_1$ are randomly selected from 0 to $\pi$ and from 0 to 2$\pi$, and 
$E_{10}$ and $E_{20}$ are randomly selected from 0 to 0.5, leading to determined 
$k_1(\lambda,\phi)$ and $k_2(\lambda,\phi)$. Here, the sampled range of $E_{10}$ and $E_{20}$ 
not only covers differences in spectral index of the emitting sources, but also accounts 
for a wide range of physical factors inherent to observing a sample of different orbiting 
systems. Then, the artificial $f(\lambda)$ can be obtained by equations above. Fourth, 
through the least-squares minimization technique, a power law function 
$A(\frac{\lambda}{5100})^{\alpha_A}$ applied to describe the artificial $f(\lambda)$ can 
lead to the determined spectral index $\alpha_A$ and the continuum luminosity $L_{A}$ at 
5100\AA. Finally, repeating the steps above 20000 times, a sample of 20000 artificial 
$f(\lambda)$ can be created.

	The dependence of $L_{A}$ on $\alpha_A$ is also shown in top panel of Fig.~\ref{sed} 
for the 20000 artificial $f(\lambda)$, apparently leading to the dependence of $L_A$ on 
$\alpha_A$ moving to the upper right. Actually, we have checked effects of different values 
of $E_{10}$ and $E_{20}$ and different periodic functions applied to describe $E(B-V,\phi)$, 
leading to similar results, but with more or less extended areas with lower $L_A$. Therefore, 
\obj~ having unique variability in $\alpha_{5100}$ as an outlier having opposite behaviour in 
the space of $L_{5100}$ versus $\alpha_{5100}$ can be explained by an assumed sub-pc BBH, 
mainly due to \obj~ well covered by the space of $L_{A}$ versus $\alpha_A$.

	Before ending the section, five points should be noted. First, if the time duration 
4487 days from MJD=52141 to 56628 of the two SDSS spectra of \obj~ is exactly an integer 
multiple of the orbital period of the assumed sub-pc BBH, the discussions above cannot be 
accepted, due to the same spatial structures for the two BH accreting systems at the two epochs. 
Therefore, it is necessary to check the orbital period in \obj~ related to an assumed sub-pc 
BBH. Second, through the procedure above with $\phi$ randomly from 0 to 6$\pi$, periodic 
variations of spectral index should be expected, such as the shown examples in bottom panel 
of Fig.~\ref{sed} (the example in purple with $\alpha_{5100}$ varying from -1.78 to -0.04, 
larger than the varying range in $\alpha_{5100}$ in \obj), indicating a new method for 
detecting sub-pc BBHs in BLAGN through periodic variation in spectral index. Third, for 
assumed sub-pc BBHs, Fig.~\ref{spec} shows a solution to the continuum emissions with 
MJD=56628 in \obj, with detailed descriptions in Appendix A. Fourth, we have assumed continuum 
emissions from central two BH accreting systems to be similar as those of individual normal 
quasars. Actually, if considering modifications related to BBHs, further evidence can be 
found to better support our conclusions, as detailed descriptions in Appendix B. Fifth, in 
order to provide further clues on importance of periodic obscurations on our conclusions, 
constant extinctions with $E_1(B-V,\phi)=E_2(B-V,\phi)=E_0$ have been considered in the same 
procedure above, as detailed descriptions in the Appendix C, to further confirm periodic 
obscurations related to orbital motions as key roles for unique variability properties of 
optical spectral index in BLAGN if harbouring sub-pc BBHs.

\begin{figure*}
\centering\includegraphics[width = 18cm,height=3.75cm]{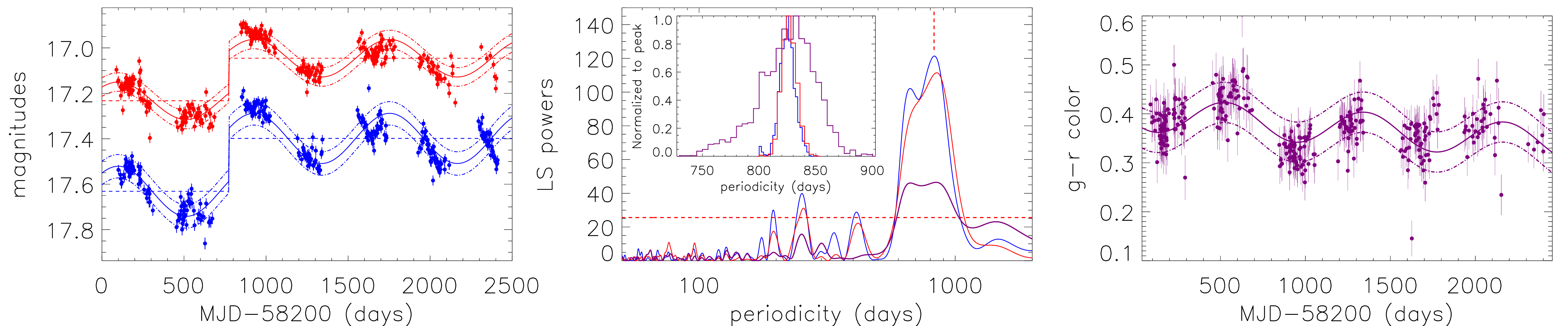}
\caption{Left panel shows the ZTF $gr$-band (in blue and in red) light curves of \obj. Solid 
and dot-dashed lines in blue and in red represent the best fitting results and the 
corresponding 1RMS scatters for the $gr$-band light curves, horizontal dashed lines in blue 
and in red show the determined base line components in the $gr$-band light curves. Middle 
panel shows the determined LS powers of the $gr$-band (in blue and in red) light curves and 
the g-r color (in purple), with horizontal dashed red line representing the 5$\sigma$ 
significance level (false alarm probability 3e-7) and with vertical dashed red line marking 
periodicity=827 days. Top left corner in middle panel shows the bootstrap method determined 
distributions (in blue, in red and in purple) of the 2000 re-determined periodicities through 
the ZTF $gr$-band light curves, and through the g-r color. Right panel shows the g-r color, 
with solid and dot-dashed lines in purple as the best fitting results and the corresponding 
1RMS scatters.}
\label{lmc}
\end{figure*}

\section{Optical QPOs in \obj} 

	The 6.3 years-long ZTF $gr$-band light curves are shown in left panel of Fig.~\ref{lmc} 
with $t=MJD-58200$ from 88 to 2411. Based on applications of a base line plus a sine component 
\begin{equation}
LC(t)~=~A\times\sin(\frac{2\pi\times t}{T_p}~+~\phi_0) + 
	\left\{
		\begin{aligned}
			B_0 \ \ \ \ (t<800) \\
			C_0 \ \ \ \ (t>800) \\
		\end{aligned}
		\right.
\end{equation}
through the least-squares minimization technique, the best fitting results with 
$\chi^2/dof\sim5.5$ to the $gr$-band light curves can lead to the determined optical periodicity 
827$\pm$2 days. Here, due to about 0.2-0.25 magnitudes darker in the $LC$ with $t<800$ than in 
the $LC$ with $t>800$, two different base lines $B_0$ and $C_0$ are applied for the $LC$ with 
$t<800$ and with $t>800$. 

	Meanwhile, the commonly accepted Lomb-Scargle (LS) periodogram \citep{zk09, vj18} has 
been applied to the ZTF $gr$-band light curves of \obj~ after subtractions of the base line 
components, leading to apparent periodicity around 820 days with significance levels higher 
than 5$\sigma$ as shown in middle panel of Fig.~\ref{lmc}. Then, the bootstrap method within 
2000 loops is applied to determine the periodicity distributions. For each loop, about half of 
the data points are randomly selected from the original $gr$-band light curves, leading to the 
LS re-determined periodicities through the re-created light curves. Distributions of the 2000 
re-determined periodicities shown in top left corner of middle panel of Fig.~\ref{lmc} lead 
to the determined optical periodicity 825$\pm$6 days and 826$\pm$6 days through the $gr$-band 
light curves.

	Furthermore, as discussed in \citet{vu16} and in our recent works in \citet{zh23a, 
zh23b, lz25, zh25b}, there are apparent effects of red noise (caused by stochastic AGN 
variability) on detecting optical QPOs. Here, a simple procedure is applied to check the 
effects of red noise on optical QPOs in \obj. Accepted the Continuous AutoRegressive (CAR) 
process described in \citet{kbs09} to describe red noise, probability of detecting fake 
QPOs can be estimated through a series of simulated light curves. The CAR process parameters 
are the intrinsic variability timescale $\tau$ and amplitude $\sigma_*$. As well discussed 
in \citet{koz10} for normal quasars, the $\tau/{\rm days}$ can be randomly selected from 
100 to 1500, and $\sigma_*/(\rm mag/days^{0.5})$ can be randomly selected from 0.003 to 
0.03. Then, 100000 artificial light curves with the same time information of the ZTF 
g-band light curve of \obj~ can be created by the CAR process with mean magnitude about 
17.5 mags (the mean value of ZTF g-band light curve of \obj). Then, the LS periodogram 
is applied to detect optical QPOs with significance level higher than 5$\sigma$, and with 
periodicity within the range from 800 to 850 days, leading to 140 artificial light curves 
to be selected. Therefore, the probability is only about 0.14\% (140/1d5) to detect fake 
optical QPOs through the CAR process simulated artificial light curves. In other words, 
the confidence level should be at least higher than 3.2$\sigma$ to support the detected 
optical QPOs in \obj~ not from red noise.

	Therefore, the optical QPOs with periodicity 827$\pm$6 days can be accepted in \obj, 
leading the time duration 4487 days from MJD=52141 to 56628 to be $5.43_{-0.05}^{+0.03}$ times 
of the periodicity. Therefore, for an assumed sub-pc BBH with different phases, there are 
different spatial structures for the two BH accreting systems at the two epochs in \obj, 
supporting the discussions on an assumed sub-pc BBH to explain the unique variability of 
spectral indices in \obj. 

	Furthermore, properties of g-r color shown in right panel of Fig.~\ref{lmc} are simply 
discussed, because the g-r color can be simply applied to trace optical spectral index. The 
corresponding LS powers of the g-r color shown in the middle panel of Fig.~\ref{lmc} can lead 
to the determined periodicity 828$\pm$23 days, with uncertainty determined by the bootstrap 
method. Meanwhile, as shown in right panel of Fig.~\ref{lmc}, the g-r color can be well modeled 
by a sine function with periodicity 819$\pm$19 days plus a linear trend, leading to the best 
fitting results with $\chi^2/dof\sim1.07$. Meanwhile, through the F-test technique  
applied to compare a sine plus linear model with a linear model, the sine component is 
preferred with confidence level higher than 10$\sigma$ as described in Appendix D. The 
periodicity determined in the g-r color is consistent with the ones determined in the $gr$-band 
light curves. Therefore, the QPOs in the g-r color can be accepted as further clues to support 
our conclusions above, and to further support a central sub-pc BBH in \obj.

\section{Conclusions}

	Considering time dependent periodic obscurations of central BH accreting systems in 
sub-pc BBHs, unique variability properties of optical continuum emissions are considered as 
clues to support central sub-pc BBHs in BLAGN. Through the oversimplified simulations, 
variability in $\alpha_{5100}$ in multi-epoch spectra leading BLAGN to be outliers in the space 
of $\alpha_{5100}$ versus $L_{5100}$ could be accepted as clues to support central sub-pc BBHs, 
such as the results in \obj~ which has unique variability pattern for dependence of 
$\alpha_{5100}$ on $L_{5100}$ very different from that for normal SDSS quasars. Meanwhile, 
periodic variations of optical spectral index could be expected in BLAGN harboring assumed 
sub-pc BBHs. The results in this manuscript open a new window to detect sub-pc BBHs by 
applications of spectroscopic properties of optical continuum emissions, as well as commonly 
applied optical QPOs.

\begin{acknowledgements}
Zhang gratefully acknowledge the anonymous referee for giving us constructive comments and 
suggestions to greatly improve the paper. Zhang gratefully thanks the kind financial support 
from GuangXi University and the kind grant support from NSFC-12173020 and NSFC-12373014 and 
the support from Guangxi Talent Programme (Highland of Innovation Talents). This manuscript 
has made use of the data from SDSS (\url{https://www.sdss.org/}) and ZTF 
(\url{https://www.ztf.caltech.edu/}).
\end{acknowledgements}

\appendix
\section{Two components related to an assumed sub-pc BBH for continuum emissions with MJD=56628}

	Besides the results shown in top panel of Fig.~\ref{sed}, an extremely oversimplified 
model has been applied to explain the variability of optical continuum emissions of SDSS 
J0012-1022. Accepted a central sub-pc BBH in SDSS J0012-1022, for the continuum emissions 
with MJD=52141, contribution related to one central BH accreting system (the first system) 
can be obtained without effects of obscurations, but no contribution related to the other 
one central BH accreting system (the second system) can be obtained due to the second BH 
accreting system totally obscured by the first system. Then, for the continuum emissions 
with MJD=56628\ in SDSS J0012-1022, due to orbital rotations, continuum emissions from both 
the central two BH accreting systems can be obtained, but with apparent obscurations on the 
continuum emissions from the first system which is partly obscured by the second system, 
and no obscurations on the continuum emissions from the second system which is now the 
foreground system. Then, even considering the extremely simple conditions, a lot of solutions 
can be applied to explain the continuum emissions with MJD=56628 in SDSS J0012-1022. Here, 
accepted E(B-V)$\sim$0.1 for the obscurations on the first system, the continuum emissions 
from the second system can be determined as
\begin{equation}
	f_{\lambda}~=~(32.51\pm0.07)\times(\frac{\lambda}{5100})^{0.32\pm0.02},
\end{equation}
as shown dashed dark green line in Fig.~\ref{spec}. Meanwhile, based on the corresponding 
continuum luminosity $\log(L_{5100}/{\rm erg/s})=44.4$ and spectral index 0.32, the estimated 
continuum emissions from the second system are similar as the ones in normal quasar SDSS 
J105635+414602 (plate-mjd-fiberid=1362-53050-0107, $\log(L_{5100}/{\rm erg/s})=44.46$ and 
$\alpha_{5100}=0.32$) listed in the database of \citet{sr11}. In one word, assumed a central 
sub-pc BBH in SDSS J0012-1022, unique variability properties of optical spectral index can 
be explained in SDSS J0012-1022.

\section{Modifications to continuum emissions related to BBHs}

	As recently discussed in \citet{ps21}, relative to effective temperature $T_{(eff,O)}$ 
from standard accretion disks for normal BLAGN, after considering effects of perturbation of 
the accretion disc of one BH accreting system due to the gravitational interaction caused by 
the companion and considering effects of spiral streams accumulating additional materials in 
each BH accreting system, modified effective temperatures $T_{(eff,i)}$ for the central two 
BH accreting systems in BBHs can be given as
\begin{equation}
	T_{(eff, i)}~\sim~T_{(eff, O, i)}(1 + q\frac{R_i}{r})^{1/4}(f_{E,i})^{1/4} \ \ \ (i=1,2)
\end{equation}
with $q$ as BH mass ratio, and $r$ as the distance between the two BH accreting systems, and 
$R$ as the radius of the part of the perturbed disk component to the central BH of the $i$th 
BH accreting system, and $f_{E,i}$ as the ratio of accretion rate to Eddington accretion rate. 
Therefore, relative to $T_{(eff,O)}$, the modified $T_{(eff, i)}$ are commonly larger, leading 
to stronger continuum emissions but with common spectral indices. Therefore, considering 
modified effective temperatures $T_{(eff,i)}$ can lead to the simulated dependence $L_{A}$ 
on $\alpha_A$ moving to the upper right with longer distances in top panel of Fig.~\ref{sed}, 
to further support our conclusions.

\section{Effects of periodic obscurations on dependence of $L_{5100}$ versus $\alpha_{5100}$}

\begin{figure}
\centering\includegraphics[width = 9cm,height=6cm]{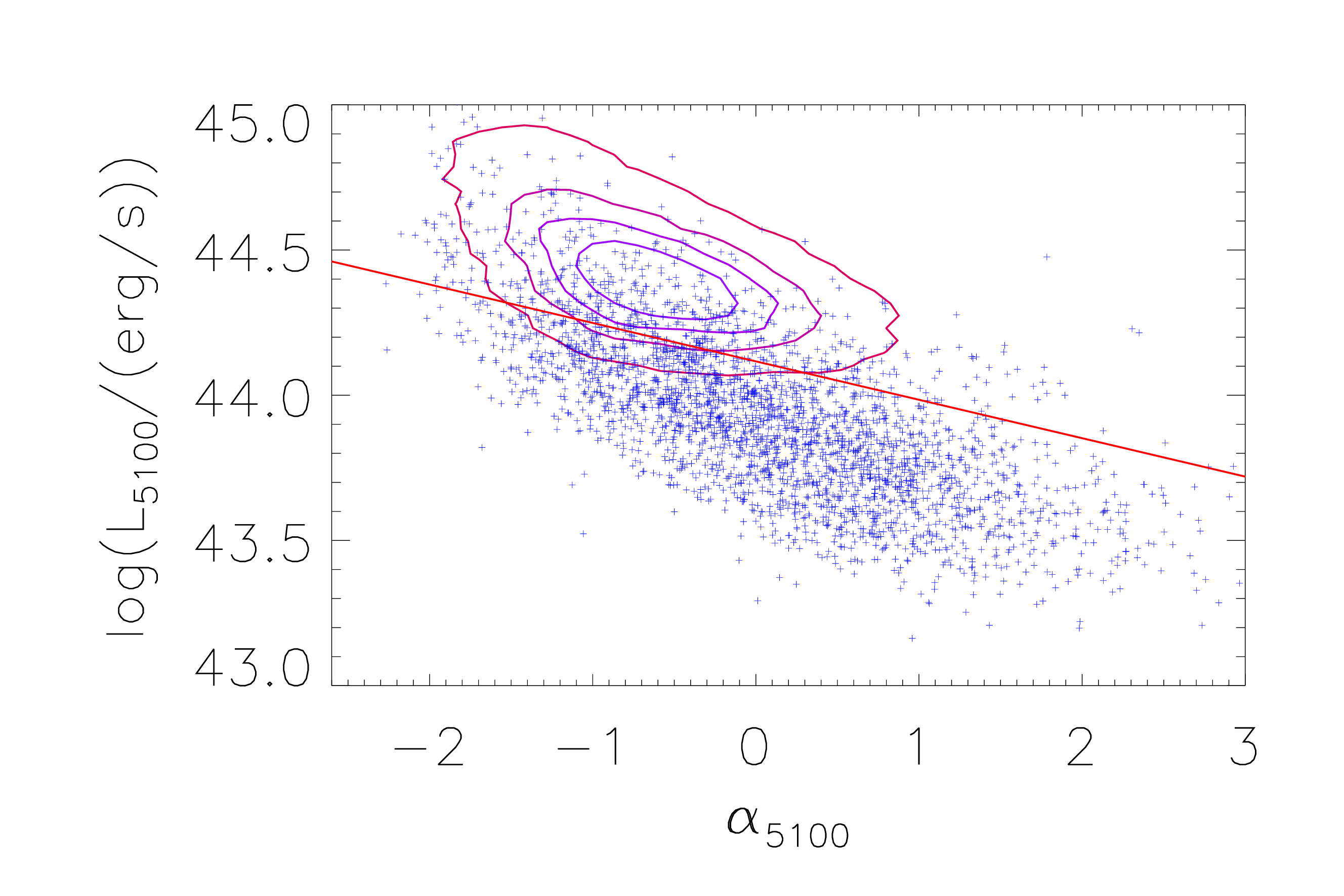}
\caption{Contour with reddish levels represents the same results shown in top panel of 
Fig.~\ref{sed} after considering periodic obscurations in assumed sub-pc BBHs. Pluses in blue 
represent the results from the 3000 newly created artificial $f_\lambda$ after considering 
$E_1(B-V,\phi)~=~E_2(B-V,\phi)~=~E_0$ with $E_0$ as a random value between 0 and 0.5. The 
solid red line shows the dependence determined from SDSS quasars as shown in top panel of 
Fig.~\ref{sed}.}
\label{s2ed}
\end{figure}

	Besides the simulated results shown in top panel of Fig.~\ref{sed}, it is necessary 
to check whether the periodic obscurations have key roles to determine the unique variability 
of optical spectral index in BLAGN if harbouring central sub-pc BBHS. Here, accepted that 
the same procedure applied but with a constant extinction by considering 
$E_1(B-V,\phi)=E_2(B-V,\phi)=E_0$ with $E_0$ as a random value between 0 and 0.5, the 
corresponding results are shown in Fig.~\ref{s2ed} based on newly created 3000 artificial 
$f_\lambda$. It can be found that the new results cannot totally cover the same regions of 
the results with considerations of periodic obscurations, but have more extended areas to 
the bottom right corner in the space of $L_{5100}$ versus $\alpha_{5100}$. Therefore, 
periodic obscurations related to orbital motions have key roles for unique variability 
properties of optical spectral index in BLAGN if harbouring sub-pc BBHs.

\section{F-test results on the sine component in the g-r color}

	Besides the best fitting results with $\chi_1^2 = 345.663$ and $dof_1=322$ by a sine 
component plus a linear trend shown in right panel of Fig.~\ref{lmc}, only a linear trend has 
also been applied, leading to $\chi_2^2 = 492.948$ and $dof_2=325$. Therefore, through the 
F-test technique, the calculated value $F_p$ is about 
\begin{equation}
F_p = \frac{\frac{\chi_2^2 - \chi_2^1}{dof_2-dof_1}}{\chi_1^2/dof_1}=45.73
\end{equation}. 
Meanwhile, considering $dof_2-dof_1=3$ and $dof_1=322$ as the number of $dof$s of the F 
distribution numerator and denominator, the F distribution expected value with 10$\sigma$ 
confidence level is about 20 which is smaller than $F_p=45.73$. Therefore, the sine component 
included in the g-r color is preferred with confidence level higher than 10$\sigma$.

\end{document}